\documentclass[sigconf,natbib=false]{acmart}
\usepackage{subcaption}
\usepackage{multirow}
\usepackage{caption}
\usepackage{cprotect}

\AtBeginDocument{%
  }

\setcopyright{acmcopyright}
\copyrightyear{2023}
\acmYear{2023}
\acmDOI{XXXXXXX.XXXXXXX}

\RequirePackage[
  datamodel=acmdatamodel,
  style=acmnumeric,
  ]{biblatex}

\begin{document}

\title[FrameFinder: Explorative Multi-Perspective Framing Extraction from News Headlines]{FrameFinder: Explorative Multi-Perspective\protect\\ Framing Extraction from News Headlines}

\author{Markus Reiter-Haas}
\email{reiter-haas@tugraz.at}
\orcid{0000-0001-9852-8206}
\affiliation{%
  TU Graz
 \country{Austria}
}

\author{Beate Kl\"{o}sch}
\email{beate.kloesch@uni-graz.at}
\orcid{0000-0002-8061-6088}
\affiliation{%
 University of Graz
 \country{Austria}
}

\author{Markus Hadler}
\email{markus.hadler@uni-graz.at}
\orcid{0000-0002-0359-5789}
\affiliation{%
 University of Graz
 \country{Austria}
}

\author{Elisabeth Lex}
\email{elisabeth.lex@tugraz.at}
\orcid{0000-0001-5293-2967}
\affiliation{%
  TU Graz
 \country{Austria}
}

\renewcommand{\shortauthors}{Reiter-Haas et al.}

\begin{abstract}
Revealing the framing of news articles is an important yet neglected task in information seeking and retrieval. In the present work, we present FrameFinder, an open tool for extracting and analyzing frames in textual data. FrameFinder visually represents the frames of text from three perspectives, i.e., (i) frame labels, (ii) frame dimensions, and (iii) frame structure. By analyzing the well-established gun violence frame corpus, we demonstrate the merits of our proposed solution to support social science research and call for subsequent integration into information interactions.
\end{abstract}



\def\UrlFont{\ttfamily\small} %
\maketitle

\section{Introduction}

Cognitive biases, such as framing effects, influence information seeking and retrieval behaviors~\cite{azzopardi2021cognitive}. In this vein, biased search results have been shown to affect user attitudes due to exposure~\cite{draws2021not}. Moreover, it has been well established in psychology that framing also affects the behavior and choices of people~\cite{tversky1981framing}.
Detecting and understanding the framing of online news is thus important due to its influence on readers, but also very challenging~\cite{morstatter2018identifying}. While there are several approaches for computational framing analysis~(see \cite{ali2022survey} for an overview), many rely on annotated data and train a classifier. However, framing is defined as \emph{the selection and salience of aspects in a communicating text}~\cite{entman1993framing} and thus requires a deeper understanding than just doing predictions. Moreover, even in such supervised settings, the amount of available data is typically rather sparse. The sparsity issue was therefore one of the main challenges in the recent shared framing detection task at SemEval 2023~\cite{piskorski2023semeval}, where only a few or even zero samples were available per language. Notably, the best performing teams all used pretrained Transformers to tackle the task at hand~\cite{reiter2023mcpt,liao2023marseclipse,wu2023sheffieldveraai}. Due to these challenges, the landscape of framing detection tools is still shallow, especially regarding openly available ones (e.g., \cite{bhatia2021openframing}).

In the present work, we expand upon existing computational framing research by providing a novel tool to discover and extract frames from texts with a focus on online news. \emph{FrameFinder} extracts frames from three distinct perspectives using Transformer models~\cite{vaswani2017attention}. As described in \cite{reiter2023exploration}, frames can be analyzed (i) by their associated \emph{frame labels}, (ii) their \emph{frame dimensions}, and (iii) their \emph{frame structure}. To showcase the benefits of the tool, we conducted an analysis on the gun violence frame corpus (GVFC)~\cite{liu2019detecting}. There we find that the discussion is mostly framed regarding security rather than health, despite the names of involved people being a major structural element. Besides, the \textbf{openly available library and online demonstration}\cprotect\footnote{The demo is available at: \textbf{\url{https://huggingface.co/spaces/Iseratho/frame-finder}} and accompanied by a brief video introduction: \textbf{\url{https://iseratho.github.io/external/frame-finder-video.html}} The underlying code is also available as a standalone Python library for full customization of algorithms and configuration: \textbf{\url{https://github.com/Iseratho/framefinder}} that can be installed via \emph{pip}: \verb|pip install framefinder|}
allows both \underline{social science researchers} and novice users to analyze the framing of texts without requiring technical (e.g., programming) skills. For future research, we strive to incorporate framing analyses directly into the retrieval process of online news to accomplish more balanced media consumption of users, either by informing them about the framing bias or by adapting, e.g., reranking, the retrieved results.

\setlength{\textfloatsep}{6pt}

\begin{figure}[t]
    \centering
    \includegraphics[scale=0.7]{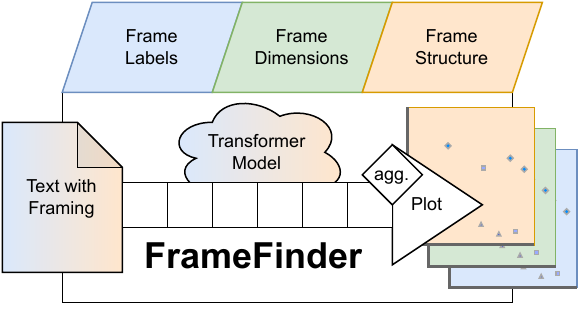}
    \caption{Schematic overview of the framing detection tool.}
    \label{fig:overview}
\end{figure}

\section{FrameFinder: Framing Detection}
\emph{Framing} has multiple definitions across various scientific disciplines~\cite{sullivan2023three}. In this work, we consider communicative frames following Entman~\cite{entman1993framing} regarding the \emph{selection and salience of aspects in a communicating text to promote a specific interpretation}. As a result, framing deals with the presentation on both a micro and macro level~\cite{scheufele2007framing}. Due to such nuances, framing is difficult to identify for algorithms~\cite{morstatter2018identifying}. Therefore, conceptualizations of framing are often only partially considered in automatic text processing~\cite{ali2022survey}.

\emph{FrameFinder} is a tool to discover and extract frames from textual data using multiple distinct perspectives.
As depicted in Figure~\ref{fig:overview}, the tool takes texts as input that are (deliberately or undeliberately) framed in a certain way. The aim is to extract those frames in a human-comprehensible manner. To that end, we use the expressive power of Transformer models~\cite{vaswani2017attention}. Internally, Transformers use embeddings, i.e., numerical vectors, to create rich representations of texts and parts thereof. The output representations, which can be probability vectors, alignment scores, or graph representations, are then aggregated and plotted. As previously identified in~\cite{reiter2023exploration}, we consider three distinct types of representations for framing analysis, i.e., frame labels, frame dimensions, and frame structure. For each type of representation, FrameFinder aggregates the result (when analyzing more than 1 sample) and visualizes them in a suitable format.
Taken together, such a multi-perspective view of the data allows for a more nuanced framing analysis and the customizability of the library enables an explorative way to not only detect established but also discover novel frames. 

\begin{figure}[t]
    \centering
    \begin{tabular}{c}
        \includegraphics[scale=.45]{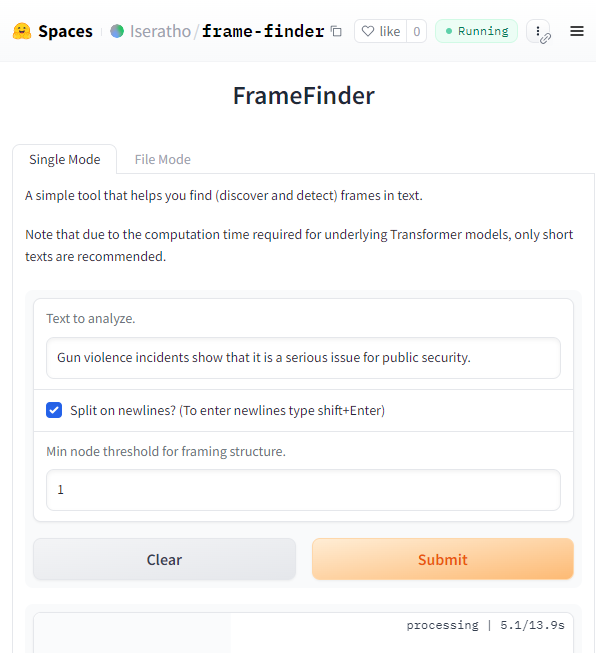}
    \end{tabular}
    \caption{Truncated Screenshot of the Online Demo. For an overview of the generated plots, refer to experimental results in Figure~\ref{fig:demo}.}
    \label{fig:screen}
\end{figure}

\paragraph{Online Demonstration.}
For the online demonstration, we built the core part using HuggingFace Transformers~\cite{wolf2019huggingface} library and models. Together with Gradio~\cite{abid2019gradio}, we deployed it as a HuggingFace space (see Figure~\ref{fig:screen}).
The demo runs on a CPU-only instance with 16 GB of RAM. 
The basic interface comprises two modes, a text-based and a file-based mode. 
The first allows entering example(s) in a text box, while the latter requires the upload of a text file. In both modes, the text is by default split on newlines into individual documents that are analyzed and aggregated. This option can be disabled to analyze the corpus as a single document (which is only recommended for short texts, as both the probability of frames being present and text structure tend to increase with text length).
Additionally, there is a filtering option for the structural visualization based on node occurrence within graphs (i.e., the degree-weighted frequency across individual graphs). Finally, in the text-based mode, a few examples are provided that are cached (i.e., pre-computed) and thus evaluated instantly.
For the deployed configuration (i.e., models and definitions of labels/dimensions) refer to the detailed description in Section~\ref{sec:gvfc_demo} that was conducted with the same settings.

In the following, we describe the basic approaches of the three types of framing perspectives. Afterward, we discuss the relation of the tool to social science research.

\subsection{Frame Labels}
Framing detection can be approached as a classification task, in which specific \emph{frame labels} are predicted to be either present or absent. This typically requires an annotated corpus. However, such corpora are scarce, with notable examples including the media frame corpus~\cite{card2015media}, the gun violence frame corpus~\cite{liu2019detecting}, and the SemEval 2023 Task 3 Subtask 2 corpus~\cite{piskorski2023semeval}. Moreover, the number of samples within these corpora are typically rather small\footnote{The media frame corpus version 2 contains three subcorpora with $6327$ on average but is deprecated due to changes in LexisNexis interface. The gun violence frame corpus contains $2990$ samples, while the shared task in SemEval contains $2,049$ split among train/dev/test set and nine languages (with three languages only in test).}. Alternatively, when given label definitions, the label prediction can also be modeled as a zero-shot prediction task. 

Recent efforts to predict frame labels include contributions to the SemEval tasks (e.g., \cite{wu2023sheffieldveraai,liao2023marseclipse,reiter2023mcpt}) and the OpenFraming tool~\cite{bhatia2021openframing}. The latter differentiates between \emph{frame discovery} using topic models and \emph{frame prediction}, which involves training a classification model. Due to this explorative nature, it is similar in spirit to FrameFinder but requires expert knowledge and labor to annotate the data through content analysis. In contrast, we strive to avoid manual annotations, by considering multiple perspectives instead.

For aggregation of the prediction, we consider the mean and standard error of the label probabilities per sample. We then visualize the aggregated scores using a bar chart, and typically consider a threshold of $0.5$ (denoted by color) to be indicative of which frame labels to assign to the corpus as a whole. 

\subsection{Frame Dimensions}

Some frames are defined antagonistically, such as concerning moral foundations~\cite{haidt2004intuitive}. Considering the antagonistic care/harm pair, a text can be framed either positively emphasizing care (i.e., as a virtue) or negatively with harm in mind (i.e., as a vice), but not both. Such dimensions can be analyzed by considering the alignment within the embedding spaces of words and documents. 
Example approaches of dimensional framing analysis are moral framing in news~\cite{mokhberian2020moral}, political framing on social media~\cite{jing2021characterizing}, or both, i.e., moral framing of political messages on social media~\cite{reiter2021studying}.

The framing of documents can be analyzed either on a per-word basis using e.g. Word2Vec~\cite{mikolov2013efficient} or on a per-document basis using e.g. Sentence-Transformers~\cite{reimers2019sentence}. In both scenarios, the position of an embedding (of a word or document) concerning the anchor embeddings (from the antagonistic pair) is determined. Herein, the FrameAxis method~\cite{kwak2021frameaxis} scores the \emph{frame bias} and intensity by projecting embeddings onto the axis formed by the antagonistic pair. The frame bias is defined as the mean of the scores, while the intensity considers the variance. Hence, the former specifies the leaning towards a frame, while the latter determines the activity along an axis. In the present work, we use FrameAxis for aggregating alignment scores but apply it to documents rather than words. The dimensions are plotted using horizontal lines, with the position of the projected points specifying the bias and their size specifying the intensity after aggregation.

\subsection{Frame Structure}

Some frames within a text are even more nuanced and require the consideration of the semantic structure. In this regard, the relations between the parts of text (e.g., words or phrases) are vital to extract the framing. One potential method for structural analysis is semantic role labeling (SRL)~\cite{gildea2002automatic} that assigns tags that identify the type of argument in relation to a predicate. Two common examples of semantic roles are the \emph{agent} tag, which is typically the subject, and \emph{patients}, which are usually objects. An example approach for framing analysis is detailed in \cite{jing2021characterizing}, where the agents and patients are visualized as tree stumps. 

Alternatively, abstract meaning representations (AMR)~\cite{banarescu2013abstract} explicitly capture the semantic relations as rooted, directed, acyclic graphs\footnote{For details of the node and edge types refer to the guidelines: \url{https://github.com/amrisi/amr-guidelines/blob/master/amr.md}}. In addition to extracting the semantic roles, these semantic graphs transform words and phrases into simplified semantic concepts, which improves comparability and subsequent transformations. Therefore, and in line with~\cite{reiter2023exploration}, we use AMR in the present work. When aggregating multiple semantic graphs, we create a weighted metagraph by superimposition of individual graphs. Thus, more pronounced concepts and relations get more emphasis, while additionally allowing filtering operations to only retain the most common elements of the metagraph.

\subsection{Relation to Social Science Research}

In the social sciences, such as sociology or communication studies, the analysis of frames also plays an important role, particularly in qualitative social research such as content analysis. Here, texts are typically coded manually, either deductively, i.e., on the basis of predetermined theoretical aspects~\cite{mayring2015}, or inductively derived from the data material, as in the case of grounded theory~\cite{strübing2014}. FrameFinder works similarly to deductive content analysis by assigning pre-defined frames, i.e., frame labels (2.1) or moral dimensions (2.2), as codes to text passages. The detection of frame structures (2.3) is comparable to the basic principles of axial coding in the grounded theory approach, where identified codes and concepts, i.e., frames, are interpretatively contrasted and linked to each other.
A tool like FrameFinder can help to get a first impression of the frames used in the text corpora and to decide on the further way of analysis. The frames found can then be integrated into MAXQDA~\cite{verbi2021maxqda} or other qualitative coding software for more in-depth analysis.
However, social researchers need to consider the pre-defined labels and dimensions that underlie this tool in order to interpret and extend their manual analyses accordingly. The adoption of frame detection tools such as FrameFinder in social science research will depend on the choice of underlying framing concepts and their adaptability to various contexts and research goals.

\begin{figure*}[t]
    \centering
    \begin{minipage}{0.44\linewidth}%
    \begin{subfigure}{0.9\linewidth}
		\includegraphics[width=\linewidth]{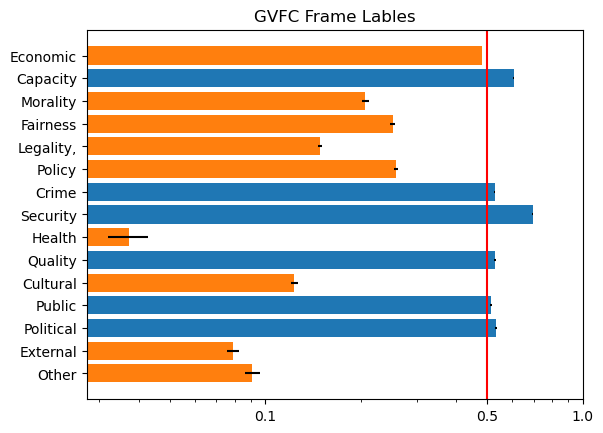}
		\caption{Frame Labels}
		\label{fig:labels}
	\end{subfigure}
    \begin{subfigure}{0.9\linewidth}
		\includegraphics[width=\linewidth]{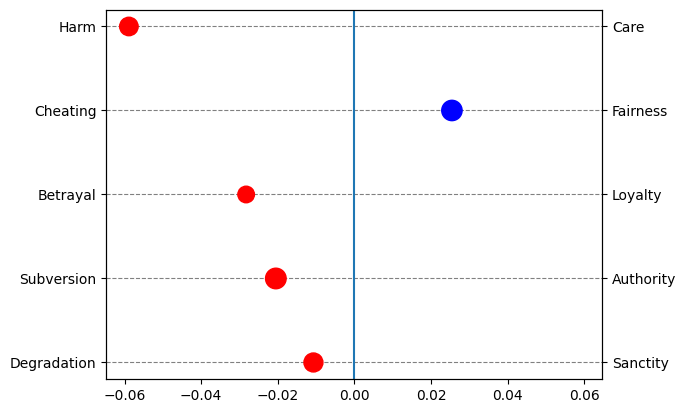}
		\caption{Frame Dimensions}
		\label{fig:dimension}
	\end{subfigure}
    \end{minipage}%
    \hspace{0.5mm}
    \begin{minipage}{0.55\linewidth}%
        \begin{subfigure}{0.99\linewidth}
        \includegraphics[width=\linewidth]{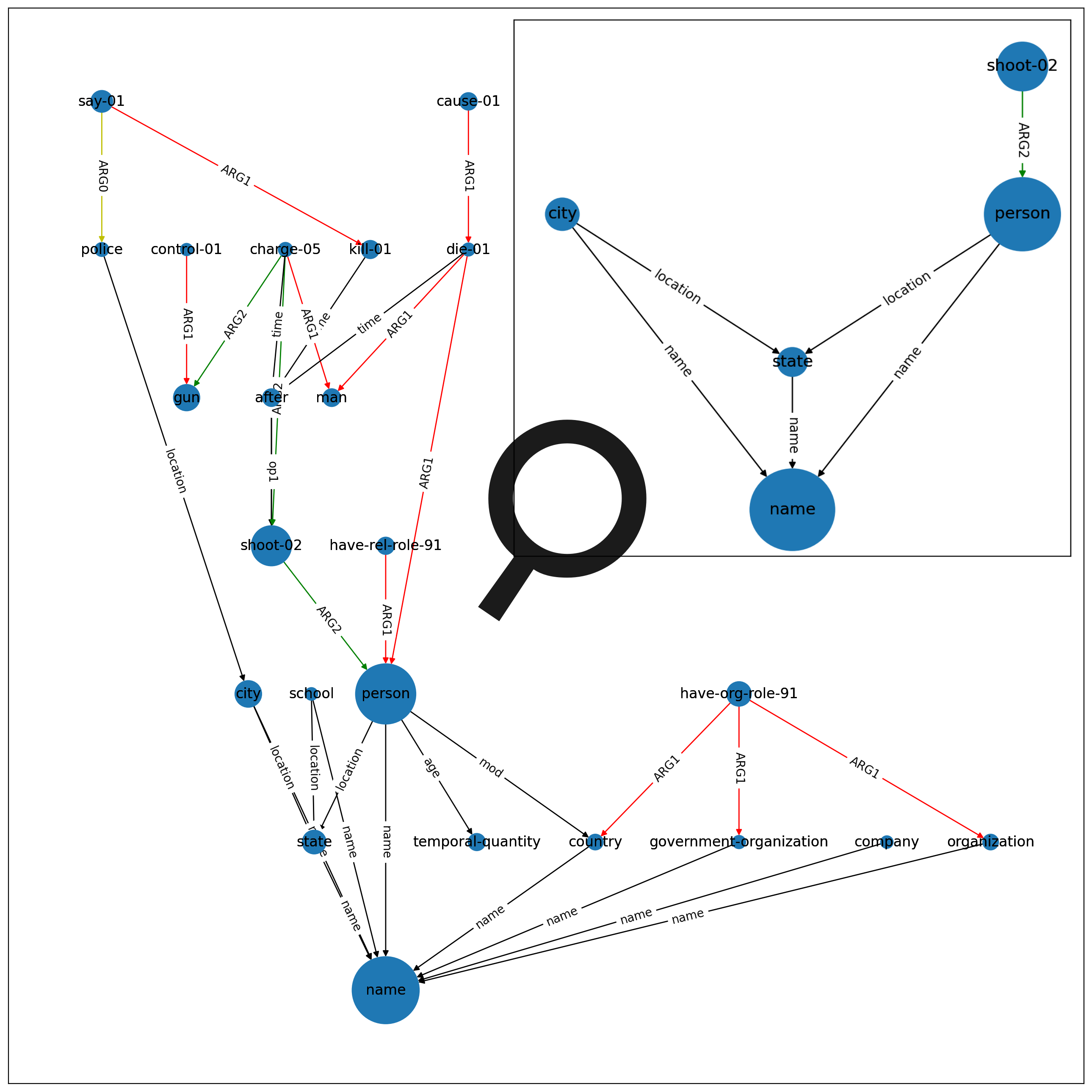}
		\caption{Frame Structure (with zoomed-in substructure at the top right).}
		\label{fig:structure}
	\end{subfigure}
    \end{minipage}%

    \caption{Framing visualizations of the GVFC. }
    \label{fig:demo}
\end{figure*}

\begin{table}[t]
    \centering
    \begin{tabular}{lrr}
    \toprule
    \textbf{GVFC Themes} & \# Events & \# Issues \\
    \midrule
    Total headlines & 1269 & 1339 \\
    \midrule
    Economic consequences & 3 & 92 \\
    Gun control/regulation & 16 & 306 \\
    Gun/2nd Amendment rights & 7 & 59 \\
    Mental health & 51 & 29 \\
    Politics & 32 & 401 \\
    Public opinion & 18 & 244 \\
    Race/ethnicity & 84 & 50 \\
    School or public space safety & 28 & 156 \\
    Society/culture & 4 & 44 \\
    \midrule
    Total labels & 243 & 1381 \\
    \bottomrule
    \end{tabular}
        \caption{Statistics of the annotated GVFC.}
    \label{tab:gvfc}
    \vspace{-4mm}%
\end{table}

\section{Demonstration with the GVFC}
\label{sec:gvfc_demo}

To demonstrate the merits of the framing extraction tool, we analyze the gun violence frame corpus (GVFC)~\cite{liu2019detecting}. The corpus consists of $2990$ news headlines about gun violence in the United States. Figure~\ref{fig:demo} shows the results extracted with FrameFinder\footnote{Note that while the results were computed using the same underlying code, for efficiency, we extracted the frames using a GPU rather than using the free CPU-only online demo interface.}.

\paragraph{Models and Configuration.}
In the code, the models and their configuration can be adapted before computation. For the analysis of the GVFC, we use the same configuration (i.e., definitions of labels and dimensions), as well as models that are deployed in the online demonstration for consistency’s sake. We choose three popular models, together with the well-established labels of the media frame corpus as labels and moral foundation theory as dimensions.

For the frame label extraction, we use a zero-shot classification model based on BART~\cite{lewis2019bart}, i.e., \emph{facebook/bart-large-mnli}. For zero-shot labels, we used the $14$ specific media frames (and 1 unspecific other category) defined by their keyword list in~\cite{card2015media}.

For the frame dimensions, we use an encoder model based on MPNet~\cite{song2020mpnet}, i.e., \emph{sentence-transformers/all-mpnet-base-v2}. For the poles of the dimensions, we use the instructions from the moral foundation dictionary~\cite{frimer2017moral} (i.e., version two) of the five axes: harm/care, cheating/fairness, betrayal/loyalty, subversion/authority, and degradation/sanctity.

For the frame structure, we use another BART-based based model trained on abstract meaning representations (AMR)~\cite{banarescu2013abstract}, i.e., \emph{model\allowbreak\_parse\allowbreak\_xfm\allowbreak\_bart\allowbreak\_base-v0\_1\_0}\footnote{The model can be downloaded from the AMRlib model GitHub repo: \url{https://github.com/bjascob/amrlib-models}}. We set the threshold for nodes to $300$ and only plot the largest weakly connected component together with another zoomed-in version using a threshold of $1000$.

\paragraph{Framing Analysis.} From Figure~\ref{fig:labels}, we observe that gun violence headlines are mostly framed from a security viewpoint. Other important frames are about resources (i.e., capacity), crime, quality of life, public opinion, as well as political frames. In comparison, it is not seen as a health issue. Interestingly, there appears to be an absence of certain frames regarding morality and fairness, which can be investigated with the framing dimensions.

From a moral standpoint (see Figure~\ref{fig:dimension}), it revolves most about the harm caused. Overall, the moral framing is rather negative with betrayal, subversion, and degradation residing on the vice side. In contrast, the fairness frame is the only positive (i.e., virtue) frame invoked in the headlines. While the bias of the frames differs noticeably (i.e., regarding their positions), the differences in intensity (i.e., point size) are much less pronounced. In this regard, the subversion/authority axis appears to be more emphasized compared to the betrayal/loyalty axis. While we clearly observe differences, with the overall negativity and fairness being less biased compared to harm, it shows that these moral values are of lesser concern when framing the news headlines.

Considering the structural view of the arguments (i.e., Figure~\ref{fig:structure}) shows that, while complex in nature, the headlines have a common theme. Specifically, as shown in the zoomed-in version, the headlines typically refer to the name of the victim of the shooting rather than the shooter (which is specified by the ARG2 role). Noteworthy is that guns and police have a subordinate role in the headlines.

To summarize, gun violence headlines frame the topic as a security issue that causes harm, with specific persons, such as the victims (mentioned by their names), being a focal point.

\paragraph{Comparison to GVFC Annotations.}
Here, we compare our results with the ground truth labels of the GVFC. In GVFC, headlines can either be assigned to singular events/incidents or issues of gun violence as an ongoing problem. Additionally, each headline gets assigned zero to two labels that determine the theme of the news story. We provide an aggregated overview in Table~\ref{tab:gvfc} (refer to \cite{liu2019detecting} for further details). Both types (i.e., events and issues) appear roughly equally, but issues are far more often associated with labels.%

This highlights a limitation of non-exploratory framing analysis, which involves first creating a codebook and then applying it to a corpus. Our use of FrameFinder reveals that the corpus often emphasizes the victims in event headlines. Similar to the annotations, we observe that politics and public opinion are common themes, while mental health gets neglected. In sum, while the annotations and findings from the exploratory framing analysis using FrameFinder largely align, the latter offers additional insights, e.g., emphasis on victims, which was not explicitly annotated in the corpus.

\section{Conclusion}

Framing analysis is intrinsically explorative and spans multiple disciplines. To advance research in this complex field, we present FrameFinder: an \emph{explorative multi-perspective framing extraction} tool. Our user-friendly online demo offers insights into three distinct types of framing present in a text.

Currently, FrameFinder is designed to serve as a support tool for social science researchers. However, we recommend extending its application to information retrieval systems in future work. With media biases being a societal concern~\cite{hamborg2019automated}, we advocate for the development of more refined automatic models for media analysis.

\printbibliography

\end{document}